\documentclass[showkeys,amsmath,amssymb,aps,preprint,groupedaddress,superscriptaddress]{revtex4-1}
\usepackage{amsmath}    
\usepackage{amsfonts}  
\usepackage[latin1]{inputenc}

\begin{document}
\newcommand{\scalarriemann}[1]{\stackrel{#1}{R}}
\newcommand{\einstein}[2]{\stackrel{#1}{G}^{\mathrm{\raisebox{-0.1cm}{$_{#2}$}}}}

\title{An Extension of Teleparallelism and the Geometrization of the Electromagnetic Field}
\author{J. B. Formiga}
\email{jansen.formiga@uespi.br}   
 \affiliation{Centro de Ciências da Natureza, Universidade Estadual do Piauí, C. Postal 381, 64002-150 Teresina, Piauí, Brazil}

\author{J. B. Fonseca--Neto}
\email{jfonseca@fisica.ufpb.br}

\author{C. Romero}
\email{cromero@fisica.ufpb.br}

\affiliation{Universidade Federal da Paraíba, Departamento de Física, C. Postal 5008, 58051-970 João Pessoa, Pb, Brazil}
\date{\today}

\begin{abstract}
As is well known, both Weyl and Weitzenb\"ock spacetimes were initially used as attempts to geometrize the electromagnetic field. In this letter, we prove that this field can also be regarded as a geometrical quantity in an extended version of the Weitzenb\"ock spacetime. The new geometry encompasses features of both Weyl  and Weitzenb\"ock spacetimes.  In addition, we obtain Einstein's field equations coupled to the Maxwell energy-momentum tensor from a purely geometrical action and, to exemplify the advantage of using this new geometry when dealing with conformal invariance, we construct a model that is equivalent to a known conformal invariant teleparallel model.
\end{abstract}


\maketitle

\section{Introduction}
In order to geometrize the electromagnetism, many famous physicists have spent a great amount of their time generalizing the geometrical framework upon which general relativity is founded. Among them are names such as Weyl, Kaluza and Einstein. While Kaluza generalized this framework by adding an extra dimension \cite{Kaluza:1921tu},  Weyl and Einstein took completely different approaches. In Weyl's approach, a nonmetricity tensor known as Weyl $1$-form was added to the spacetime manifold \cite{weyl1952space}. Einstein, in turn, considered a kind of geometry (Weitzenb\"ock spacetime) where gravity is described by torsion, and not by curvature as in general relativity (GR) \cite{Sauer:2004hj}. However, it seems that all these attempts did not succeed in providing a satisfactory geometrical representation of the electromagnetic field \cite{Goenner:2005xs,lrr-2004-2}.

Nowadays, Weyl geometry and Weitzenb\"ock spacetime are still important geometries because of their richness. For instance, the theory formulated by Einstein in the framework of Weitzenb\"ock spacetime, which is known as teleparallelism, is used for  solving the problem of the localization of the gravitational energy \cite{Maluf:2005kn,aldrovandi2012teleparallel}. With respect to Weyl geometry, we might say that one of its most important features is that it provides a natural setting for conformal invariance \cite{Hammond:2002rm}.

The main goal of this paper is to show that, by extending teleparallelism theory to Weyl geometry, one is able to construct a geometrical action that is equivalent to the Einstein-Hilbert one plus the electromagnetic action in curved spacetime, which yields Einstein's field equations with the Maxwell energy-momentum tensor. We will call this kind of theory ``Weyl Teleparallel theory'' (WTT). It is also shown that the WTTs allow the introduction of conformal invariance in a much easier way than teleparallel theories do. In doing so, we show an equivalence between the conformally invariant teleparallel theory of Ref. \cite{Maluf:2011kf} and a particular WTT.

This article is organized as follows. In Sec. 2, we set the notation and convention used in this paper, as well as   the basic mathematical facts of Weyl geometry. We proceed to Sec. 3 to briefly introduce teleparallelism theory. All the results are left to Sec. \ref{s16012013a}, where the WTT is presented.

\section{Notation and conventions } \label{s160102013c}
Throughout this paper the holonomic and the anholonomic indices are denoted by Greek and Latin letters, respectively. The tetrad fields are represented by $e_A$ (frame) and $e^A$ (coframe), whose components in the coordinate basis are denoted by $e_A^{\ \ \mu}$ and $e^A_{\ \ \mu}$, respectively; the coordinate basis is denoted by $\partial_{\mu}$. The components of the metric tensor in the tetrad basis are $\eta_{AB}=diag(+1,-1,-1,-1)$, while the ones in the coordinate basis are $g_{\mu \nu}$. We use square brackets around indices to represent the antisymmetric part of a tensor.

Let $M$ be a manifold endowed with a metric $g$ and a linear connection $\nabla$. In this paper, the definition of torsion, curvature and the Weyl nonmetricity condition are given by 
\begin{eqnarray}
T(V,U)\equiv \nabla_V U-\nabla_U V-[V,U], \label{14012013a}\\
R(V,U)W\equiv \nabla_V\nabla_U W-\nabla_U\nabla_V W-\nabla_{[V,U]} W, \label{14012013b}\\
\sigma(V)g=\nabla_V g, \label{14012013c}
\end{eqnarray}
where $\sigma$ is the Weyl $1$-form, and $V,U,W$ are vectors belonging to the tangent bundle of $M$. Unless stated otherwise, the components of these tensors are defined as $T^A_{\ \ BC}\equiv <e^A, T(e_B,e_C)>$ and $R^A_{\ \ DBC}\equiv <e^A,R(e_B,e_C)e_D>$ .

To keep Eq.~(\ref{14012013c}) invariant under the conformal transformation $\tilde{g}=e^{2\theta}g$, where $\theta$ is a scalar function, one demands that 
\begin{equation}
\tilde{\sigma}=\sigma+2d\theta, \label{17012013b}
\end{equation}
where $d$ is the exterior derivative operator.

\section{Teleparallel theories} \label{s21012013a}
\subsection{The geometrical setting} 
Many different geometries can be specified by setting  one or more quantities in the definitions (\ref{14012013a})-(\ref{14012013c}) equal to zero. For instance, if we set $T=\sigma=0$, we have the Riemannian geometry. In turn, if we set $R=\sigma=0$, we obtain the Weitzenb\"ock spacetime. The latter corresponds to the geometry in which the teleparallel theories are formulated.

In teleparallel theories, one assumes the existence of a particular tetrad $e_A$ that satisfies
\begin{equation}
\nabla_{e_B}e_A=0. \label{14012013d}
\end{equation}
This is equivalent to saying that there exists a basis in which the affine connection coefficients vanish. Of course, in a Riemannian manifold, this would imply that $e_A$ is a holonomic basis (see Eq. (\ref{14012013a})). However, this need not be the case for a more general manifold. In the case of teleparallelism, one removes this restriction by assuming a nonvanishing torsion tensor. 

Substitution of (\ref{14012013d}) into (\ref{14012013a})-(\ref{14012013c}) yields
\begin{eqnarray}
T(e_A,e_B)=-[e_A,e_B],\\
R(V,U)W=0,\\
\sigma=0.
\end{eqnarray}

In general, teleparallel theories are based upon the following general Lagrangian density
\begin{eqnarray}
{\cal L}_{T}=e\bigl(  a_1Q^{ABC}Q_{BAC}+a_2Q^AQ_A
\nonumber \\
+a_3Q^{ABC}Q_{ABC} \bigr),
\end{eqnarray}
where 
\begin{equation}
Q^A_{\ \ BC}\equiv <e^A,T(e_B,e_C)>=2e_B^{\ \ \mu}e_C^{\ \ \nu} e^A_{\ \ [\nu,\, \mu]} \label{9012013b}
\end{equation}
 are the components of the Weitzenb\"ock connection in the preferred frame $e_A$, the comma stands for the partial derivative, and we have defined $Q_A\equiv Q^B_{\ \ B A}$, and $e\equiv det(e^A_{\ \ \mu})$. 

For $a_1=-1/2$, $a_2=1$ and $a_3=-1/4$, we have the TEGR (teleparallel equivalent of general relativity) \cite{aldrovandi2012teleparallel}. As the name suggests, the TEGR is formally equivalent to GR. 

\subsection{Teleparallel theories with conformal invariance } \label{s160102013d}

In Ref. \cite{Maluf:2011kf}, the authors consider a teleparallel model that is invariant under the transformation
\begin{equation}
\tilde{g}=e^{2\theta}g,\quad \tilde{e}^A=e^{\theta}e^A,\quad \tilde{e}_A=e^{-\theta}e_A, \label{8012013f}
\end{equation}
where the tilde indicates a new basis and $\theta$ is a function of the coordinates; it is easy to verify that $\tilde{e}=e^{4\theta}e$. 

The Lagrangian density of this model is given by
\begin{eqnarray}
{\cal L}_T= e \biggl[ \phi^2 \biggl( -\frac{1}{4}Q^{ABC}Q_{ABC}-\frac{1}{2}Q^{ABC}Q_{BAC}
\nonumber \\
 +\frac{1}{3}Q^AQ_A \biggr)+6g^{\mu \nu} \phi_{|\mu}\phi_{|\nu} \biggr], \label{9012013a}
\end{eqnarray}
where $\phi$ is a scalar field that is assumed to transform as 
$\tilde{\phi}=e^{-\theta}\phi$ under (\ref{8012013f}). In addition, it is also defined {\it a gauge covariant derivative} whose components are $\phi_{|\mu}\equiv \left( \partial_{\mu}-Q_{\mu}/3 \right)\phi $. 

The Lagrangian density (\ref{9012013a}) is invariant under (\ref{8012013f}). In fact, any term like
\begin{equation}
L=a_1Q^{ABC}Q_{BAC}+a_2Q^AQ_A+a_3Q^{ABC}Q_{ABC} \label{9012013c}
\end{equation}
with 
\begin{equation}
a_1+3a_2+2a_3=0  \label{17012013a}
\end{equation}
will transform as $\tilde{L}=e^{-2\theta}L$, which can be used as a start point to construct many different conformal invariant theories. As we shall see, in the WTT we can start from terms that are simpler than (\ref{9012013c}).

\section{Weyl teleparallel theories (WTTs) }\label{s16012013a}

Let $M$ be a manifold endowed with a metric $g$, a connection $\nabla$ and a $1$-form $\sigma$. Now suppose there exists a privileged frame $\{e_A \}$ that satisfies
\begin{equation}
\nabla_{e_B}e_A=-\frac{1}{2} \sigma_B e_A \label{8012013a}
\end{equation}
It is clear that any other frame related to $\{ e_A\}$ by a constant Lorentz transformation will also satisfy this condition. 

By using Eq. (\ref{8012013a}) in the definitions (\ref{14012013a})-(\ref{14012013c}), one obtains
\begin{eqnarray}
T^A_{\  \ BC}=2e_B^{\ \ \mu}e_C^{\ \ \nu} e^A_{\ \ [\nu,\, \mu]} +\sigma_{[C|}\delta^A_{\ \ |B]}, \label{8012013b} \\
R^A_{\ \ DBC}=e_B^{\ \ \alpha}e_C^{\ \ \mu} \sigma_{[\alpha,\, \mu]}\delta^A_{\ \ D}, \label{8012013c}
\end{eqnarray}
while Eq. (\ref{14012013c}) is satisfied identically. Note that the curvature vanishes for an integrable Weyl geometry.

It is straightforward to show that Eqs. (\ref{14012013b}) and (\ref{8012013c}) lead to
\begin{eqnarray}
e\scalarriemann{c}\doteq e\biggl(-\frac{1}{4}T^{ABC}T_{ABC}-\frac{1}{2}T^{ABC}T_{BAC} +T^AT_A
\nonumber \\
+\frac{3}{2}\sigma_A\sigma^A-2\sigma_AT^A \biggr), \label{8012013d}
\end{eqnarray}
where $\scalarriemann{c}$ is the scalar curvature in terms of the Christoffel symbols, all surface terms have been neglected, and $T_A\equiv T^B_{\ \ B A}$.

\subsection{The electromagnetic field as a geometric entity \label{ss19012013a}}

Let us consider the following action:
\begin{eqnarray}
S=\int d^4x e \biggl(4R^{AB}R_{AB}-\frac{1}{4}T^{ABC}T_{ABC}
\nonumber \\
-\frac{1}{2}T^{ABC}T_{BAC} +T^AT_A   +\frac{3}{2}\sigma_A\sigma^A-2\sigma_AT^A \biggr), \label{8012013e}
\end{eqnarray}
where we are using relativistic units. In order to obtain the field equations, one may vary $S$ with respect to the tetrad and the Weyl field independently or, equivalently, take the metric and the Weyl field as independent variables. 

By identifying  $R_{AB}$ with $F_{AB}/2$, where $F_{AB}$ is the electromagnetic tensor, and using the identity (\ref{8012013d}) in the action (\ref{8012013e}), one arrives at the Einstein-Hilbert action minimally coupled with the electromagnetic field (see, e.g., pp. 153 and 163 of Ref. \cite{Inverno}). Therefore, Einstein's field equations with the Maxwell energy-momentum tensor follow naturally.  However, it should be noted here that we have a clear difference between the two approaches: in the case of WTT we can readily see the geometric nature of the electromagnetic field as it can be naturally identified with the Weyl field. It is also important to note that the derivation by purely geometrical means of the Einstein field equations with the Maxwell energy-momentum tensor as source  is not a result exclusive of this model (see, e.g., Refs.  \cite{Poplawski:2008mq,Poplawski:2008pq,citeulike:5303504}).

\subsection{Equation of motion}
Let us now set $\sigma=0$ (no electromagnetic field). If we couple a matter field with (\ref{8012013e}) and vary the action with respect to the metric, we will clearly obtain Einstein's field equations
\begin{equation}
\einstein{c}{\mu\nu}=8\pi T^{\mu\nu},
\end{equation}
where $\einstein{c}{\mu\nu}$ is the Einstein tensor written in terms of the Christoffel symbols, and $T^{\mu\nu}$ is the energy-momentum tensor. Since $\einstein{c}{\mu\nu}_{\ \ :\mu}=0$, where the colon represents the Riemannian covariant derivative, we have $T^{\mu\nu}_{\ \ : \mu}=0$. It can be verified that this last result leads to the geodesic equation with the Christoffel symbols, as in GR (see, e.g., p. 152 of Ref. \cite{papapetrou1974lectures}).

\subsection{Conformal invariance \label{s16012013b}}
To introduce the conformal invariance in the teleparallel model (\ref{9012013a}), one needed to postulate a scalar field which is not present in the original geometry (Riemann-Cartan) and add some extra properties to it. It is possible to get rid of this scalar field by taking terms like $eLL'$, where $L'$ is written in the same fashion as $L$ [see Eqs.~(\ref{9012013c}) and (\ref{17012013a})]. However, the resultant theory would probably be too complicated and we would still be restricted by the conditions (\ref{17012013a}). Here, we show that a natural conformal invariance can be achieved in the case of WTT with an integrable Weyl field playing the role of $\phi$. 

For an integrable Weyl field we have $\sigma=\varphi_{,\mu}dx^{\mu}$, where $\varphi$ is a scalar function. In this case, the transformation (\ref{8012013f})  leads Eq. (\ref{17012013b})  to
\begin{equation}
\tilde{\varphi}=\varphi+2\theta.
\end{equation}
From  (\ref{8012013b}), it is straightforward to verify that
\begin{eqnarray}
\tilde{T}^A_{\  \ BC}=e^{-\theta}T^A_{\  \ BC}, \\
\tilde{T}_A=e^{-\theta}T_A.
\end{eqnarray}
 It is interesting to note that, since $\eta_{AB}$ does not change, we can raise and lower tetrad indices without changing these transformations.

From the Lagrangian density
\begin{eqnarray}
{\cal L}_I=e e^{-\varphi}\bigl(  a_1T^{ABC}T_{BAC}+a_2T^AT_A
\nonumber \\
+a_3T^{ABC}T_{ABC}    \bigr), \label{3022013a}
\end{eqnarray}
it is easy to build up many conformal invariant theories regardless of the values of $a_i$ (i=1,2,3). When one imposes the condition (\ref{17012013a}), the terms with $\varphi$ in brackets in Eq. (\ref{3022013a}) cancel out.

\subsection{The WTT equivalent of (\ref{9012013a}) } \label{ss19012013b}
  By identifying $e^{-\varphi}$ with $\phi^2$ and using the relation $T_{ABC}=Q_{ABC}+\sigma_{[C}\eta_{B]A}$, one can easily check that the Lagrangian density (\ref{9012013a}) is equivalent to 
\begin{eqnarray}
{\cal L}= e e^{-\varphi} \bigl( -\frac{1}{4}T^{ABC}T_{ABC}-\frac{1}{2}T^{ABC}T_{BAC}
\nonumber \\
+T^AT_A \bigr) \label{22012013a}
\end{eqnarray}
As one can see, the Lagrangian density (\ref{22012013a}) looks like more natural than (\ref{9012013a}) because it contains only geometrical quantities.

\section{Final remarks}
In principle, the WTT presented in subsection \ref{ss19012013a} may suffer from the same problem as Weyl theory, namely, ``the second clock effect''. Since this effect was predicted by Einstein from a geometrical point of view, Weyl argued that it may not happen because the behavior of real clocks should be deduced only from a dynamical theory of matter \cite{Straumann:883608}. If Weyl's argument is right, then the behavior of clocks in this WTT may be the same as that of GR, since the field equations and the equation of motion are already the same. 

It is worth mentioning that, unlike Weyl,  we have not demanded that the theory be invariant by Weyl transformations. This demand led Weyl to a complicated theory that is not formally equivalent to GR. In this case, the model (\ref{8012013e}) may become more suitable for the geometrization of the electromagnetism.

Since the equivalence shown in subsection \ref{ss19012013b} holds with an integrable Weyl geometry, the second clock effect is not present. Therefore, this equivalence may hold not only in terms of the field equations and the equation of motion, but also in terms  of measurements.

\section*{Acknowledgments}
C. Romero would like to thank CNPq for financial support.


\end{document}